\documentstyle[emulateapj] {article}

\lefthead{EVANS ET AL.}
\righthead{CO IN IRAS 14348-1447}
 
\begin{document}

\title{Resolved CO($1\to0$) Nuclei in IRAS 14348-1447: Evidence for
Massive Bulge Progenitors to Ultraluminous Infrared Galaxies}

\author{A. S. Evans\altaffilmark{1,2},
J. A. Surace\altaffilmark{3},
\& J. M. Mazzarella\altaffilmark{3}
}
 
\altaffiltext{1}{Division of Physics, Math, \& Astronomy,
California Institute of Technology, Pasadena, CA 91125}
\altaffiltext{2}{Present Address: Dept. of Physics \& Astronomy,
SUNY, Stony Brook, NY 11794-3800: aevans@mail.astro.sunysb.edu}
\altaffiltext{3}{Infrared Processiong \& Analysis Center, California
Institute of Technology, MS 100-22, Pasadena, CA 91125: 
jason,mazz@ipac.caltech.edu}

\begin{abstract}

High-resolution, CO(1$\to$0) interferometry of the ultraluminous infrared
galaxy IRAS 14348-1447 is presented. The merger system has a molecular gas
mass of $\sim3\times10^{10}$ M$_\odot$ and a projected nuclear separation
of 4.8 kpc (3.5\arcsec), making it one of the most molecular gas-rich
galaxies known and an ideal candidate for studying the intermediate stages
of an ultraluminous merger event.  The CO morphology shows two molecular
gas components associated with the stellar nuclei of the progenitors,
consistent with the idea that the molecular disks are gravitationally
bound by the dense bulges of the progenitor galaxies as the interaction
proceeds. In contrast, less luminous infrared galaxies observed to date
with projected nuclear separations of $\lesssim 5$ kpc show a dominant CO
component between the stellar nuclei.  This discrepancy may be an
indication that the progenitors of mergers with lower infrared luminosity
do not possess massive bulges, and that the gas is stripped during the
initial encounter of their progenitors. A comparison of the CO and radio
luminosities of the NE and SW component show them to have comparable radio
and CO flux ratios of $f(NE)/f(SW) \sim0.6$, possibly indicating that the
amount of star-forming molecular gas in the progenitors is correlated with
the supernovae rate.  The estimate of molecular gas masses of the nuclei
and the extent of the radio emission are used to infer that the nuclei of
IR 14348-1447 have gas densities comparable to the cores of elliptical
galaxies.

\end{abstract}

\keywords{galaxies: ISM---infrared: galaxies---ISM: molecules---radio
lines: galaxies---galaxies: active---individual: IRAS 14348-1447}

\section{Introduction}

Galaxy tidal interactions and mergers are responsible for the most
luminous galaxy phenomenon in the universe, be they starbursts or active
galactic nuclei (AGN). The energy sources are fueled by molecular gas,
which is subject to gravitational torques, dynamical friction and
dissipation during the interaction. As a result, the distribution of young
stars in the merger and the efficiency at which an AGN can be built or
fueled depends not only on the amount of fuel available, but also on the
Hubble-types of the progenitors (e.g., Mihos \& Hernquist 1996; Mihos
1999).

The most energetic examples of mergers known are the ultraluminous
infrared galaxies (ULIGs: $L_{\rm IR} [8-1000\mu{\rm m}] \gtrsim 10^{12}$
L$_\odot$).  ULIGs have been of particular interest in the last 15 years,
both because of their extreme starburst environment (e.g. Joseph \& Wright
1985) and because of their possible evolutionary connection with QSOs and
radio galaxies (Sanders et al.  1988a,b; Mirabel, Sanders, \& Kazes 1989).

In order to determine the properties of the progenitors of ULIGs and the
mechanisms at work as the progenitors come under the gravitational
influence of each other, a program has been initiated to map the
distribution of CO($1\to0$) in several ULIGs for which the nuclei of the
progenitors have yet to coalesce.  In this {\it Letter}, high-resolution
CO($1\to0$) observations of the ULIG IRAS 14348-1447, obtained with the
Owens Valley Millimeter Array (OVRO), are presented.  With an infrared
luminosity $L_{\rm IR} = 1.8\times10^{12}$ L$_\odot$ and a molecular gas
mass of $\sim 4.2\times10^{10}$ M$_\odot$ (Sanders, Scoville, \& Soifer
1991), IRAS 14348-1447 is one of the most luminous and molecular gas rich
ULIGs that show no definitive evidence of AGN activity (i.e, broad line
emission or strong high ionization lines; see Veilleux, Sanders, \& Kim
1997).  The nuclei have a projected separation of $\sim3.5\arcsec$
(4.8 kpc: Carico et al. 1990; Surace, Sanders \& Evans 1999; Scoville
et al.  1999), making it ideally suited for the resolution of OVRO.
The data presented here show evidence contrary to observations which
have led to suggestions that molecular gas in luminous mergers as a
class is stripped and collects between the merging stellar nuclei, but
supports recent models that show the molecular disks in massive bulge
galaxy-galaxy mergers are gravitationally stabilized against stripping
(Mihos \& Hernquist 1996; Mihos 1999). An $H_0 = 75$ km s$^{-1}$
Mpc$^{-1}$ and $q_0 = 0.5$ is assumed throughout such that 1$\arcsec$
subtends $\sim 1.4$ kpc at the redshift of the galaxy ($z=0.0825$).

\subsection{Interferometric Observations}

Aperture synthesis CO($1\to0$) maps of IRAS 14348-1447 were made with the
Owens Valley Radio Observatory (OVRO) Millimeter Array during two
observing periods from 1999 February to 1999 April. The array consists of
six 10.4 m telescopes, and the longest observed baseline was 242 m. Each
telescope was configured with $120\times4$ MHz digital correlators. During
the observations, the nearby quasar [HB89] 1334-127 (5.89 Jy at 107 GHz;
B1950.0 coordinates 13:34:59.81 -12:42:09.9) was observed every 25 minutes
to monitor phase and gain variations, and 3C 273 and 3C 345 were observed
to determine the passband structure. Finally, flux calibration
observations of Neptune were obtained.

The OVRO data were reduced and calibrated using the standard Owens Valley
data reduction package MMA (Scoville et al. 1992). The data were then
exported to the mapping program DIFMAP (Shepherd, Pearson, \& Taylor
1995), and the NRAO software package AIPs was used to extract spectra.

\section{Results}

Figure 1a shows a 0.8 $\mu$m archival image of IRAS 14348-1447 taken with
the Hubble Space Telescope (HST) WFPC2 instrument.  The merger consists of
two nearly face-on colliding spirals with a projected nuclear separation of
3.4$\arcsec$ (4.8 kpc). A prominent tidal tail is visible extending
northward from the northeastern galaxy (hereafter IRAS 14348-1447NE), and
a much fainter countertail extends southward from the southwestern galaxy
(hereafter IRAS 14348-1447SW).  Knots, or super star clusters, are visible
around both nuclei, as well as along the tidal tail of IRAS 14348-1447NE.

Figure 1b shows the CO(1$\to$0) emission of IRAS 14348-1447 in contours
superposed on a three-color HST NICMOS (Near-Infrared Camera and
MultiObject Spectrometer) image of the ULIG. The CO emission consists of
two unresolved components (FWHM = $2.8\arcsec\times1.9\arcsec$ = 3.9 kpc
$\times$ 2.7 kpc) which are centered on the respective stellar nuclei
of the progenitors. The upper limit of the CO extent in each galaxy is
significantly less than the average effective diameter\footnote{Young et
al. (1995) define the effective diameter to be the diameter that encloses
70\% of the total CO emission.} of 11.5$\pm7.5$ kpc determined from
a sample of $\sim140$ nearby spiral galaxies (Young et al. 1995). The
total CO flux density of IRAS 14348-1447 is 40.8 Jy km s$^{-1}$ and the
CO luminosity is $L^{\prime}_{\rm CO} = 8.0\times10^9$ K km s$^{-1}$
pc$^2$, with $\sim42$\% and 68\% of the emission emanating from the
NE and SW components, respectively. The total CO flux density is 20\%
less than the flux density derived from the single-dish measurement of
IRAS 14348-1447 with the NRAO 12m Telescope (half power beam width,
HPBW = 55$\arcsec$:  Sanders, Scoville, \& Soifer 1991). Such a
discrepancy is likely due to flux calibration uncertainties associated
with the individual observations, which can each be as large as 15\%.
Assuming a standard ratio ($\alpha$) of CO luminosity to H$_2$ mass of
4 M$_\odot$ (K km s$^{-1}$ pc$^2$)$^{-1}$, which is similar to the value
determined for the bulk of the molecular gas in the Milky Way, the total
molecular gas mass of the pair is calculated to be $3.2\times10^{10}$
[$\alpha/4$] M$_\odot$, or 14 times the molecular gas mass of the Milky
Way.\footnote{Radford, Solomon, \& Downes (1991) have used theoretical
models to determine that $\alpha$ ranges from 2--5 M$_\odot$ (K km
s$^{-1}$ pc$^2$)$^{-1}$ for a reasonable range of temperatures and
densities. Downes \& Solomon (1998) have modeled interferometric CO
data of a sample of infrared luminous galaxies to derive an $\alpha =
0.8$. Thus, the molecular gas mass of IRAS 14348-1447 may be a low as
$6.4 \times10^{9}$ M$_\odot$.  However, note that the dynamical mass
derived using $\Delta v_{\rm FWHM} = 350$ and $300$ km s$^{-1}$, $r =
108$ and $84$ pc (see \S 3), and assuming the reasonable inclination
angle of the disk axis relative to the line-of-sight of $29^{\rm o}$
and $18^{\rm o}$ for the IRAS 14348-1447NE and 14348-1447SW disks,
respectively, yields masses consistent with the molecular gas mass
calculated with $\alpha = 4$ M$_\odot$ (K km s$^{-1}$ pc$^2$)$^{-1}$.}

Figure 1b also shows the CO(1$\to$0) emission-line spectra extracted from
both progenitors. The line profile of IRAS 14348-1447NE appears asymmetric
and blueshifted, consistent with the H$\alpha$ line profile determined
from Fabry-Perot observations (Mihos \& Bothun 1998), and has a full width
at half the maximum intensity line width of $\Delta v_{\rm FWHM} \sim 350$ km
s$^{-1}$.  IRAS 14348-1447SW has a more symmetric CO line profile with a
$\Delta v_{\rm FWHM} \sim 300$ km s$^{-1}$. The relative velocity offset
of the CO line centroids is approximately 120 km s$^{-1}$, consistent with
the offsets measured from the optical emission lines (Veilleux et al.
1995; Mihos \& Bothun 1998). Table 1 summarizes the properties discussed
above.

\section{Discussion}

Figure 1 clearly shows that IRAS 14348-1447 consists of two gas-rich
spiral galaxies that have undergone gravitational response to at least
one initial close approach. It is also clear that the molecular gas is
associated with the individual progenitor nuclei at this stage of the
merger process.  Further, the LINER (Low Ionization Nuclear Emission-line
Regions) emission-line spectra in both nuclei (Veilleux et al.  1995)
indicates that shocks from either a low ionization AGN or supernovae resulting
from massive starbursts with a total energy output of 1.8$\times10^{12}$
L$_\odot$ have commenced. All of these properties can be understood
in terms of a merger in which the molecular disks associated with each
progenitor have been stabilized against stripping by a dense, massive
stellar bulge. Given the upper limit on the gas distribution of IRAS
14348-1447 relative to local spiral galaxies, the interaction has likely
already driven the molecular gas inwards, resulting in enhanced nuclear
activity (see simulations by Mihos \& Hernquist 1996 and Mihos 1999).

Disk stabilization against stripping is also applicable to other
intermediate stage ULIGs.  Recent observations of Arp 220 have shown that
the major CO concentrations in this merger are on the individual stellar
nuclei, which have a projected separation of 0.95$\arcsec$ ($\sim 350$ pc;
Sakamoto et al. 1999). Likewise, PKS 1345+12, a ULIG with very warm,
Seyfert-like infrared colors relative to Arp 220 and IRAS 14348-1447,
shows the CO emission concentrated only on the active radio nucleus,
indicating that gas infall and AGN activity has been triggered by
interactions with the companion galaxy (Evans et al. 1999). The companion
galaxy has colors consistent with an elliptical galaxy (Surace et al.
1998), which explains its lack of detectable molecular gas.

In contrast, the morphologies of all of the lower luminosity luminous
infrared galaxies (LIGs: $L_{\rm IR} = 10^{11.0-11.99}$ L$_\odot$)
with small ($\lesssim$ 5 kpc) projected nuclear separations observed
to date are markedly different; Arp 244, VV 114, NGC 6240, and NGC 6090
have molecular gas predominantly in one component between the two nuclei
(Stanford et al.  1991; Yun et al. 1994; Tacconi et al. 1999; Bryant \&
Scoville 1999).  Tacconi et al. (1999) have speculated that the molecular
gas in NGC 6240 has been ram-pressure stripped, and that the nuclei may
later sweep up gas as the galaxy evolves into ULIGs such as Arp 220.
An alternative explanation may be that the processes affecting the gas
in LIGs differ from ULIGs because the progenitors of each luminosity
class differ.  Specifically, ULIGs may primarily be examples of equal mass
galaxy mergers with dense stellar bulges, thus the gas is driven into the
nuclear region of the respective progenitors as the merger advances. In
contrast, LIGs may be predominantly collisions of galaxies of different
masses and relatively low-density bulges (e.g., note the differences
in the near-infrared stellar morphologies of the progenitor galaxies
of NGC 6090 and VV 114:  Dinshaw et al. 1999; Scoville et al.  1999),
thus the gas is stripped from the progenitors with extreme efficiency.
As a result, many LIGs may never evolve into ULIGs.  A larger survey of
double nuclei ULIGs, consisting of those with obvious AGN signatures
and those without, is underway to investigate the ubiquity of these
results. Such a survey will also benefit from kinematic determinations
of the stellar bulge masses of the progenitors.

A comparison of the molecular gas and radio fluxes and morphologies of the
progenitors of IRAS 14348-1447 can be used to derive the gas densities
of the nuclei.  The measured IRAS 14348-1447NE to IRAS 14348-1447SW
flux density ratios of CO and the 8.44 GHz and 1.49 GHz radio emission
(i.e., Condon et al. 1990; 1991) yield values of $f(NE)/f(SW) \sim$ 0.61,
0.62, and 0.58, respectively. The implication of this result is that the
molecular gas mass of each component is related to the source of the radio
emission. This can be understood if the nuclear radio emission is due to
synchrotron emission from supernovae and if both galaxies have similar
initial mass functions.  Thus, the CO and radio flux density ratios of
the progenitors are similar because the same fraction of massive stars
are produced per unit of star-forming molecular gas. Therefore, if the
likely assumption is made that the extent of the radio emission of each
nucleus (FWHM(NE) $\sim 0.16\arcsec$ [220 pc] and FWHM(SW) $\sim 0.12$
[170 pc]) is similar to the true extent of the molecular gas (i.e.,
the supernovae fill the same volume as the gas they are formed from),
then the northeastern and southwestern nuclei have gas densities of $\sim
2.4\times10^3$ [$\alpha/4$] and $7.7\times10^3$ [$\alpha/4$] M$_\odot$
pc$^{-3}$, respectively.  Given the uncertainty in the value of $\alpha$,
the density and the velocity dispersion of the gas in each nucleus
are comparable to the stellar densities and velocity dispersions of
elliptical galaxy cores (Faber et al. 1997), supporting the likely
connection between ULIGs and  the formation of elliptical galaxy bulges
(Kormendy \& Sanders 1992).

Molecular gas-rich ULIGs in the local universe such as IRAS 14348-1447
may provide insights to the nature of massive galaxy formation in
the universe.  Figure 2 shows a plot of the logarithm of CO(1$\to$0)
luminosity of LIGs and ULIGs versus their redshift. On this plot,
IR 14348-1447 is shown as an asterix enclosed in a circle. The rise in
$L^{\prime}_{\rm CO}$ at $z < 0.04$ is simply due to the space density of
the flux-limited infrared luminous galaxy sample.  However, the leveling
off of $L^{\prime}_{\rm CO}$ beyond $z>0.07$ is a possible indication
that, due to self regulating processes, galaxies do not contain molecular
gas masses in excess of $\sim 4\times10^{10}$ [$\alpha/4$] M$_\odot$
(Evans et al.  1996; Frayer et al. 1999).  The observed flatness of
$L^{\prime}_{\rm CO}$ beyond $z\sim 0.07$ remains constant out to
redshifts of 4.7 (Frayer et al. 1999). Thus, in terms of the richness
of the interstellar medium, galaxies such as IRAS 14348-1447 appear to
be the low redshift counterparts of molecular gas-rich, high-redshift
galaxies detected over the last decade (e.g. see Frayer et al. 1999 for
a summary).  If a substantial fraction of these systems have nuclear
gas densities comparable to IRAS 14348-1447, then they as a class are
the likely progenitors of massive elliptical galaxies.

\acknowledgements
We thank the staff and postdoctoral scholars of the Owens Valley
Millimeter array for their support both during and after the observations
were obtained. ASE thanks D. Frayer and N. Trentham for useful discussion and
assistance. We also thank the referee for many useful suggestions. ASE
was supported by RF9736D and by NASA grant NAG 5-3042.  J.M.M. was
supported by the Jet Propulsion Laboratory, California Institute of
Technology, under contract with NASA.  The Owens Valley Millimeter Array
is a radio telescope facility operated by the California Institute of
Technology and is supported by NSF grants AST 93-14079 and AST 96-13717.
The NASA/ESA Hubble Space Telescope is operated by the Space Telescope
Science Institute managed by the Association of Universities for research
in Astronomy Inc. under NASA contract NAS5-26555.

 
\centerline{Figure Captions}
 
\vskip 0.3in
 
\noindent
Figure 1.a) Hubble Space Telescope WFPC2 image of IRAS 14348-1447.  b)
CO($1\to0$) contours of the merger superimposed on a three-color
composite NICMOS image (blue=1.1 $\mu$m, green = 1.6 $\mu$m, red=2.2
$\mu$m). The CO contours are plotted as 20\%, 30\%, 40\%, 50\%,
60\%, 70\%, 80\%, 90\%, and 99\% the peak flux of 0.0283 Jy/beam. The
CO emission for each progenitor is unresolved, with a beam FWHM of
2.8\arcsec$\times$1.9\arcsec$~$ at a position angle of -21.6$^{\rm o}$.
Extracted spectra of the SW and NE progenitors are also shown.  For the
images, north is up and east is to the left.

\vskip 0.2in
\noindent
Figure 2. The plot of log($L'_{\rm CO}$) versus redshift ($z$) for a
flux-limited sample ($f_{60 \mu{\rm m}} > 5.24$ Jy) of infrared luminous
galaxies and a sample of ultraluminous infrared galaxies.  The data have
been obtained from Sanders, Scoville, \& Soifer (1991) and Solomon et
al. (1997).  The data point representing IRAS 14348-1447 is encircled.

\vfill\eject

\begin{deluxetable}{lllcccc}
\pagestyle{empty}
\tablenum{1}
\tablewidth{0pt}
\tablecaption{Molecular Gas Properties of IRAS 14348-1447}
\tablehead{
\multicolumn{1}{c}{Component} &
\multicolumn{1}{c}{RA} &
\multicolumn{1}{c}{Dec} &
\multicolumn{1}{c}{z} &
\multicolumn{1}{c}{$\Delta v_{\rm FWHM}$} &
\multicolumn{1}{c}{$S_{\rm CO} \Delta v$} &
\multicolumn{1}{c}{$M({\rm H}_2)^{b}$} \nl
\multicolumn{1}{c}{} &
\multicolumn{2}{c}{(B1950.0)$^{a}$} &
\multicolumn{1}{c}{} &
\multicolumn{1}{c}{km s$^{-1}$} &
\multicolumn{1}{c}{Jy km s$^{-1}$} &
\multicolumn{1}{c}{$\times10^{10}$ M$_\odot$}\nl}
\startdata
14348-1447NE& 14:34:53.42 & -14:47:23.1 & 0.0823 & 350 & 15.6 & 1.2 \nl
14348-1447SW& 14:34:53.30 & -14:47:26.1 & 0.0827 & 300 & 25.2 & 2.0 \nl
Total       &   \nodata   &   \nodata   &  \nodata & \nodata & 40.8 & 3.2 \nl
\enddata
\tablenotetext{a}{The coordinates are taken from Condon et al. (1991).}
\tablenotetext{b}{The H$_2$ masses are derived assuming $\alpha=4$ (see text).}
\end{deluxetable}

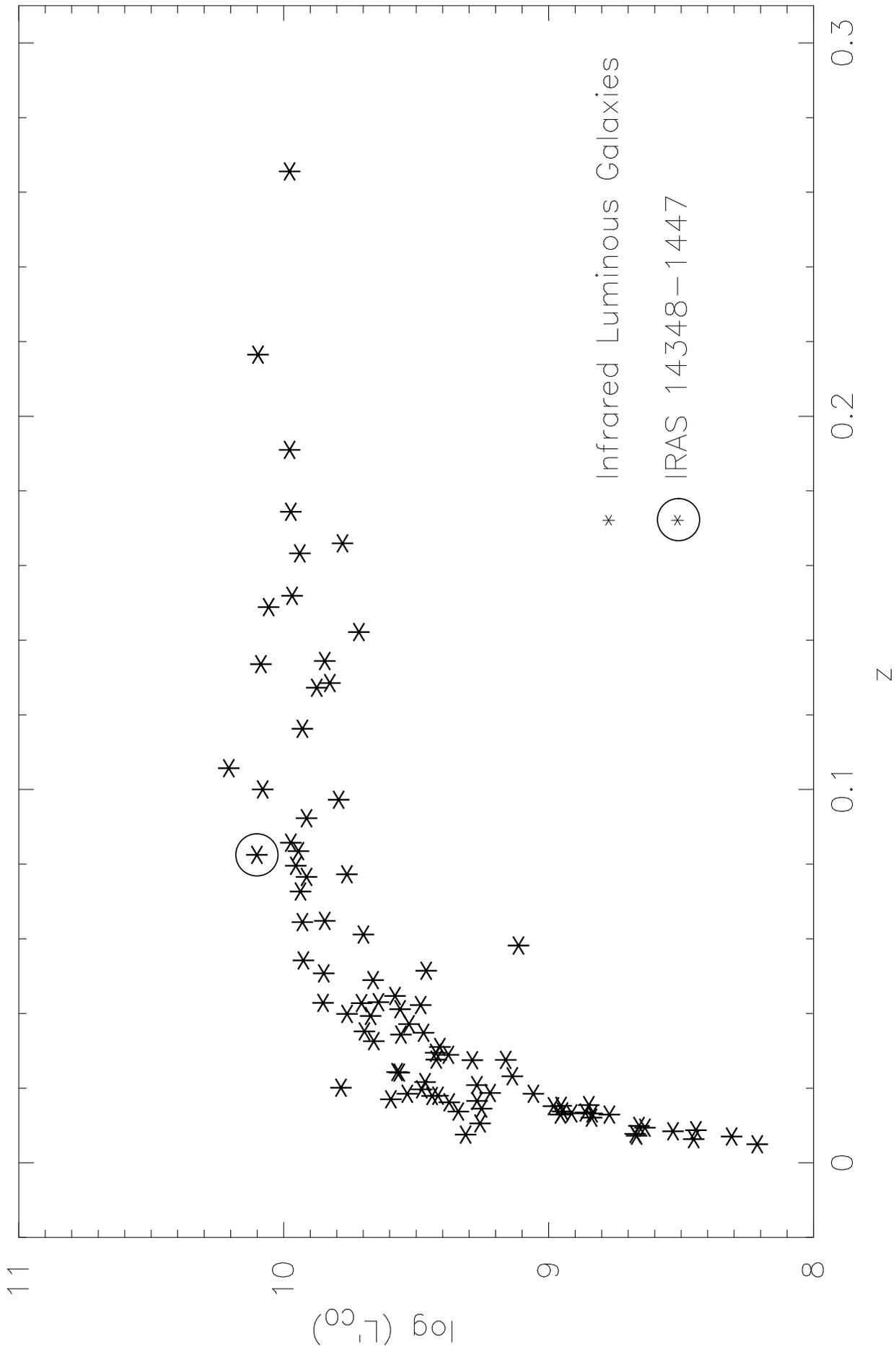
\end{document}